\documentclass[journal=jacsat,manuscript=article]{achemso}
\usepackage[utf8]{inputenc}
\usepackage[T1]{fontenc}
\usepackage[version=3]{mhchem}
\usepackage{soul}
\usepackage{bm}
\usepackage{threeparttable}
\usepackage{color}
\usepackage{multirow} 
\usepackage{array} 


\author{Qi Yu$^{1}$}
\email{qi_yu@fudan.edu.cn}
\author{Dong H. Zhang$^{1,2}$}
\author{Joel M. Bowman$^3$}
\email{jmbowma@emory.edu}

\makeatletter
\let\@fnsymbol\@arabic
\makeatother

\title{Quantum mechanical deconstruction of vibrational energy transfer rate and pathways modified by collective vibrational strong coupling}

\begin{document}
\begin{tabular}{@{}c@{}}
\textsuperscript{1}Department of Chemistry, Fudan University, Shanghai, 200438, P.R. China \\
\textsuperscript{2}State Key Laboratory of Molecular Reaction Dynamics, Dalian Institute of Chemical Physics,\\
Chinese Academy of Sciences, 457 Zhongshan Road, Dalian, 116023, P.R. China \\

\textsuperscript{3}Department of Chemistry, Emory University and Cherry L. Emerson Center \\
for Scientific Computation, Atlanta, Georgia, 30322, U.S.A. \\
\end{tabular}

\clearpage

\begin{abstract}
Recent experiments have demonstrated that vibrational strong coupling (VSC) between molecular vibrations and the optical cavity field can modify vibrational energy transfer (VET) processes in molecular systems. However, the underlying mechanisms and the behavior of individual molecules under collective VSC remain largely incomplete. In this work, we combine state-of-the-art quantum vibrational spectral calculation, quantum wavepacket dynamics simulations, and \emph{ab initio} machine-learning potential to elucidate how the vibrational dynamics of water OH stretches can be altered by VSC. Taking the \ce{(H2O)21}-cavity system as an example, we show that the collective VSC breaks the localization picture, promotes the delocalization of OH stretches, and opens new intermolecular vibrational energy pathways involving both neighboring and remote water molecules. The manipulation of the VET process relies on the alignment of the transition dipole moment orientations of the corresponding vibrational states. The emergence of new energy transfer pathways is found to be attributed to cavity-induced vibrational resonance involving OH stretches across different water molecules, along with alterations in mode coupling patterns. Our fully quantum theoretical calculations not only confirm and extend previous findings on cavity-modified energy transfer processes but also provide new insights in energy transfer processes under collective VSC.
\end{abstract}

\flushbottom

\thispagestyle{empty}
\clearpage

\section{Introduction}
Polariton chemistry has become a rapidly growing field, offering new possibilities for manipulating chemical properties by coupling molecular systems with the electromagnetic field of an optical cavity.\cite{Ebbesen2016_acr,Feist2018,Dunkelberger2022,tibben2023molecular,hirai2023molecular,simpkins2023control} In particular, the vibrational strong coupling (VSC) between molecular vibrations and the infrared cavity has garnered significant interest in physics and chemistry.\cite{Thomas2019,Ebbsen2019,Hutchison2020,ahn2023modification,Grafton2021,Xiang2020,Xiong2022science, Ebbesen2016_prl,tibben2023molecular,hirai2023molecular,simpkins2023control,Weichman2022}
It has been demonstrated that strong VSC effects, resulting in the formation of vibrational polaritons, can significantly modify chemical reactivity and selectivity.\cite{Thomas2019,Ebbsen2019,Hutchison2020,ahn2023modification,Grafton2021,Lather2019,Huo2021_jpcl,Dunkelberger2022,Schafer2022,sun2023modification,mandal2023theoretical} Experimentally, it has also been reported that VSC alters intramolecular and intermolecular vibrational energy transfer (VET) processes in model molecular systems such as \ce{W(CO)6} and \ce{Fe(CO)5}.\cite{dunkelberger2016modified,Xiang2020,Xiong2022science} 

Despite seminal experimental breakthroughs, the theoretical understanding of energy transfer processes and their underlying mechanisms remains largely elusive.\cite{mandal2023theoretical} In recent years, several theoretical studies have reported cavity-mediated VET at the single-molecule level, where only one molecule is strongly coupled to the cavity mode.\cite{Wang2022,Schafer2022,yu2023manipulating,lindoy2023quantum,schafer2024machine} While these studies provide valuable microscopic insights into VSC-modified reactions, most practical VSC experiments involve polaritonic systems with macroscopic ensembles of molecules, typically on the order of 10$^{10}$, that collectively couple to the photonic mode within the optical cavity, and the light-matter coupling strength for each molecule is relatively weak. Handling such large molecular ensembles poses significant challenges for theoretical derivations and simulations. Typically, the Tavis–Cummings model or other simplified Hamiltonians are employed to simulate VSC dynamics of many-molecule systems.\cite{Yuanzhou2019, campos2021generalization, mandal2022theory} Although these models are useful, they lack the robustness needed to capture the complexity of realistic potential energy landscapes and the influence of structural disorder.

The collective VSC effect in large molecular ensembles has also been explored through classical cavity molecular dynamics (MD) simulations, offering key insights into cavity-modified energy transfer processes.\cite{Li2020_pnas, Li2023_jacs} However, these simulations neglect quantum effects, which are essential because the formation of vibrational polaritons and their influence on chemical reactions are inherently quantum mechanical.\cite{Fregoni2022, lindoy2023quantum, Limmer2023, Richardson2023} Another persistent challenge in polariton chemistry is understanding how the dynamics of individual molecules are altered by collective VSC.\cite{sidler2020polaritonic, mandal2023theoretical, ruggenthaler2023understanding, sidler2023unraveling} Recent theoretical efforts have attempted to address this issue by simulating dilute, gas-phase molecular ensembles coupled to an optical cavity, while neglecting intermolecular interactions.\cite{sidler2020polaritonic, sidler2023unraveling} These approximations facilitate the efficient study of collective VSC effects, but they have not yet been extended to quantum dynamical simulations of realistic molecular systems coupled to cavities.

In this work, we advance beyond the single-molecule approximation by performing state-of-the-art \emph{ab initio} quantum dynamical calculations of the vibrational population dynamics of OH stretches in the \ce{(H2O)21}-cavity system, incorporating collective VSC effects. As the most common solvent and a frequent subject of VSC experiments,\cite{Lather2021, fukushima2022inherent} water’s vibrational spectra and ionic conductivity under VSC have been investigated both experimentally and theoretically.\cite{Li2020_pnas, Fukushima2021, fukushima2022inherent, lieberherr2023vibrational, Jino2024} However, the impact of VSC on water’s vibrational dynamics remains unexplored. The study of VET rates and pathways in water has long been a topic of interest,\cite{Skinner2010,Pakoulev2003,Larsen2004,Lindner2006,woutersen1999,Zhang2011,Yuki2020,Skinner2008,Hynes2009,imoto2015ultrafast,Lock2002, Hynes2004, Skinner2002, Skinner2010-2, fecko2003ultrafast, VanDerPost2015, Tokmakoff2013, Ishiyama2021} with vibrational relaxation involving a complex interplay between structural heterogeneity and the delocalized nature of OH stretches.\cite{Lock2002, Hynes2004, Skinner2002, Skinner2010-2, fecko2003ultrafast, VanDerPost2015, Tokmakoff2013, Ishiyama2021} The \ce{(H2O)21} cluster serves as a representative gas-phase model of water, capturing the essential characteristics of its hydrogen-bonding network. This system strikes a balance between computational feasibility, quantum mechanical accuracy, and the need to account for collective coupling effects in polariton chemistry. As such, \ce{(H2O)21} offers an ideal platform for investigating the quantum dynamical features of vibrational couplings and energy transfer in water under VSC.

We employ a combination of the cavity vibrational self-consistent field/configuration interaction (cav-VSCF/VCI) method\cite{Yu2022, yu2023manipulating, Yu2024JCTC} and quantum wavepacket dynamics simulations. These are integrated with our recently developed full-dimensional potential energy surfaces (PES) at the CCSD(T) level and dipole moment surfaces (DMS) at the MP2 level for water.\cite{yu2022q, Hank2015} A similar theoretical framework was previously applied to analyze the subpicosecond relaxation dynamics of the OD and OH stretches in dilute HOD-ice systems.\cite{Hank2014ab} In this study, we systematically investigate the population dynamics of the OH stretch and the associated energy transfer pathways both inside and outside the cavity, varying the light–matter coupling conditions. We further analyze and discuss the underlying mechanisms that give rise to the new landscape of vibrational dynamics under VSC. 

\section{Results}
\subsection{Vibrational spectra and population dynamics outside the cavity }
\begin{figure}[H]
\begin{center}
\includegraphics[width=0.7\textwidth]{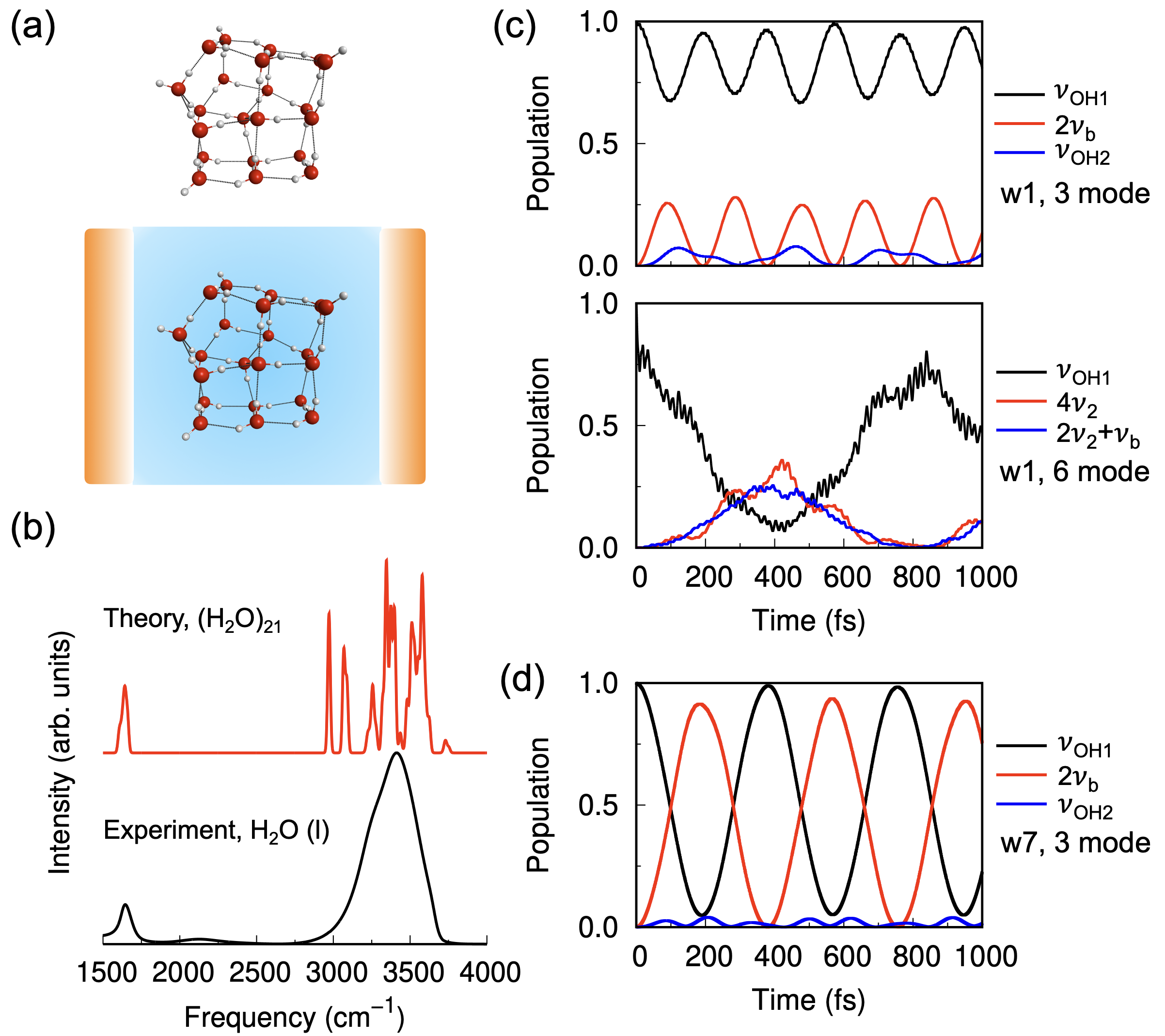}
\end{center}
\caption{\textbf{Vibrational spectra and OH population dynamics of \ce{(H2O)21} outside the optical cavity} (a) Schematic depiction of \ce{(H2O)21} placed inside and outside the optical cavity. (b) Vibrational spectra of \ce{(H2O)21} outside the cavity from anharmonic VSCF/VCI calculation and experimental IR spectrum of liquid water.\cite{bertie1996infrared} (c-d) Examples of time-dependent population of the OH stretch in monomer 1 (w1) and monomer 7 (w7) from 3 mode or 6 mode calculations. 3 mode calculation involves the bend ($\nu_{\text{b}}$) and two OH stretches ( $\nu_{\text{OH1}},\nu_{\text{OH2}}$) of one monomer. 6 mode calculation includes three additional low-frequency intermolecular modes ($\nu_1$,$\nu_2$,$\nu_3$)}
\label{fig1}
\end{figure}

Fig. \ref{fig1}a depicts the \ce{(H2O)21} cluster outside and inside the optical cavity. As detailed in Supplementary Fig. 1 and Supplementary Table 1 in Supporting Information, the 3-dimensional cage structure of \ce{(H2O)21} encompasses various H-bond topologies. The distinctive structural information can be disentangled through the vibrational spectra, and we present the simulated spectra in Fig. \ref{fig1}b. The VSCF/VCI spectrum of \ce{(H2O)21} at the optimized minimum structure, predicts the water HOH bend at approximately 1650 cm$^{-1}$ and OH stretches across a broad spectral band of 3000-3700 cm$^{-1}$ centered around 3400 cm$^{-1}$. Although we have not considered the inhomogeneous broadening by thermal effects in our VSCF/VCI calculations, our theoretical predictions align well with the experimental linear IR spectrum of liquid water \cite{bertie1996infrared}.  Overall, it can be concluded that the gas-phase \ce{(H2O)21} cluster shares a similar intermolecular H-bonding network pattern with liquid water. Note, our new cav-VSCF/VCI framework provides essentially the same spectrum of \ce{(H2O)21} as in our recent spectral work \cite{Yu2024JCTC}. This consistency arises from the fact that our new theoretical strategy focuses on the light-matter coupling terms and does not affect the molecular system when outside the cavity.  

The vibrational dynamics of OH stretch are characterized by the time-dependent change of the initial state population of excited OH stretch. Throughout the main text, we focus on two representative examples: monomer 1 (w1) and monomer 7 (w7), to illustrate the OH stretch vibrational dynamics. The complete labeling of all monomers, along with the full results for the OH stretch dynamics of all 21 monomers, is provided in the Supplementary Information. As will be shown, w1 and w7 exhibit distinct dynamic behaviors, highlighting two contrasting scenarios of vibrational relaxation. Furthermore, we define successful relaxation as a significant population decrease—falling below 0.25—within a 1000 fs timescale.

Fig. \ref{fig1}c-d provide the vibrational population dynamics of OH stretches in w1 and w7, and the full dynamics results are presented in Supplementary Fig. 2-3. First, when only three intramolecular modes of each water are considered in the VSCF/VCI and wavepacket calculations, only few successful relaxations are observed. For example, in w1, the population change of the first OH stretch displays an oscillating feature. This can be easily explained by the mild intramolecular coupling between OH stretches and HOH bend overtone. To achieve fully relaxation of the OH stretch in w1, additional low-frequency intermolecular vibrations should be included in the calculations. Indeed, we observe the relaxation of OH stretch in a separate six-mode simulation, as shown in the lower panel of Fig. \ref{fig1}c. Once intermolecular modes are considered, the population decrease of the OH stretch fundamental is strongly correlated with the rise of 4$\nu_{\text{2}}$ and 2$\nu_{\text{2}}$+$\nu_{\text{b}}$, where $\nu{_\text{2}}$ is the second low-frequency intermolecular mode included. This results from the strong mixing between OH stretch and highly excited or combination bands involving intermolecular modes. Additional examples of intermolecular mode-enabled relaxation of OH stretch are provided in Supplementary Fig. 4. The population dynamics of the OH stretch in w7, shown in Fig. \ref{fig1}d, exemplify another category of VET pathway. This intramolecular relaxation process is characterized by a significant population decrease of the excited OH stretch, accompanied by a population increase in the HOH bend overtone. Notably, an oscillatory behavior is also observed throughout the dynamics. This process is the result of the well-known Fermi resonance with the bend overtone\cite{wang2013ir} and occurs directly without the inclusion of additional intermolecular modes. This agrees with the previous work by Hynes\cite{Hynes2004} and Skinner\cite{Skinner2002,Skinner2010} in investigating the relaxation of OH stretch fundamental in liquid water. 

Our observations of the two main relaxation pathways align with the traditional localization picture. It should be noted that our theoretical set-up does not consider the vibrational couplings among intramolecular modes from different water monomers. Thus delocalization behavior of OH stretch is disregarded and direct intermolecular energy transfer among OH groups of different monomers cannot be observed. Given that vibrational relaxation occurs on a subpicosecond timescale, and the localization scheme may be a primary contributor to the inhomogeneous broadening of the spectral band, our theoretical analysis based on the local monomer approximation remains reasonable.\cite{Hank2014ab} The good agreement with previous experimental and theoretical investigations concerning spectrum and relaxation dynamics further verifies the success of our theoretical approach.\cite{Hynes2004,Skinner2010} As will be elucidated in the next section, the assumption of negligible vibrational couplings between monomers offers a clean and direct way to probe the VSC effect on the vibrational relaxations in water when optical cavity is involved. 

\subsection{VSC effects on spectra and vibrational dynamics}

\begin{figure}[H]
\begin{center}
\includegraphics[width=0.8\textwidth]{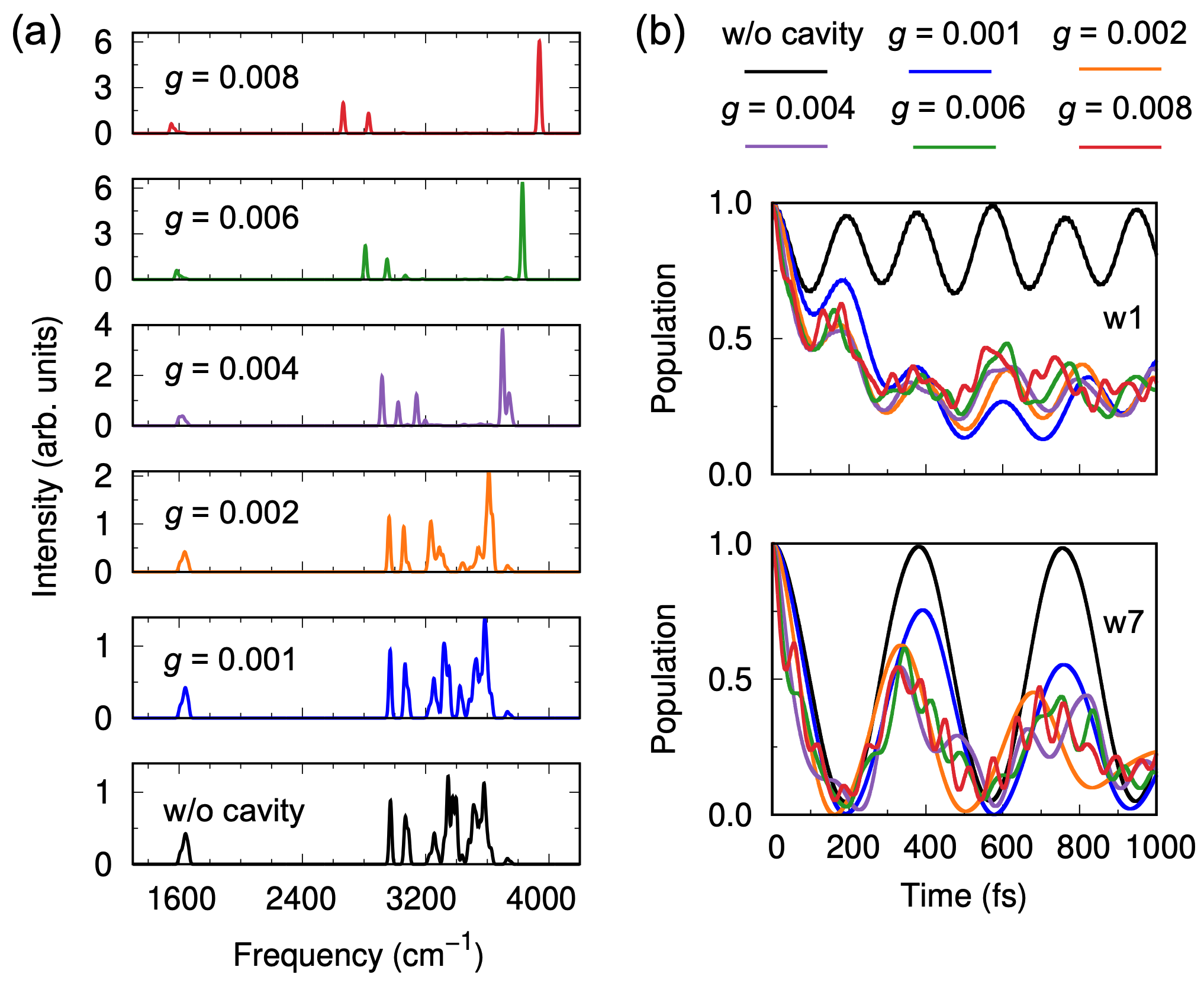}
\end{center}
\caption{\textbf{Vibrational strong coupling effects on the vibrational spectra and OH vibrational population dynamics} (a) Polaritonic vibrational spectra of \ce{(H2O)21} with different light-matter coupling strength $g$ (a.u.). (b) Examples of time-dependent population of the OH stretch in monomer 1 (w1) and monomer 7 (w7) of \ce{(H2O)21}-cavity system with different light-matter coupling factors $g$ (a.u.). The cavity frequency is set as 3400 cm$^{-1}$ for (a)-(b).}
\label{fig2}
\end{figure}

Before presenting the results on VSC effects on spectra and vibrational population dynamics, it is important to mention the differences between our theoretical setup and realistic experimental conditions. In VSC experiments, a macroscopic number of molecules, $N_{\text{mol}}\approx 10^{10}$, collectively couple to the cavity mode, and the spectral Rabi splitting scales with $\sqrt{N_{\text{mol}}/\tilde{V}}$. To compensate for the lack of $\approx 10^{10}$ molecules in our simulations, the coupling strength $g$ in this work reflects a much smaller effective cavity volume $\tilde{V}$ compared to experimental setups. For instance, with cavity frequency $\omega=3400$ cm$^{-1}$ and coupling strength $g=0.006$ a.u., the effective cavity volume $\tilde{V}$ is 0.4 nm$^3$ which falls within the picocavity regime. Nonetheless, our \ce{(H2O)21}-cavity system represents a significant step forward compared to single-molecule systems, which has been widely employed in both classical and quantum mechanical studies.\cite{Schafer2022, sun2023modification, Wang2022} Moreover, our theoretical setup and findings are also relevant to plasmonic cavities, which enable substantial single-molecule-level couplings.\cite{chikkaraddy2016single,ojambati2019quantum}

When \ce{(H2O)21} is placed inside the optical cavity with significant light-matter coupling, the direct observation of the VSC effect is the polaritonic vibrational spectra. Fig. \ref{fig2}a illustrates the vibrational spectra of the \ce{(H2O)21}-cavity system under various light-matter coupling strengths with cavity mode frequency of 3400 cm$^{-1}$. Before discussing the spectral results, it should be noted that our new cav-VSCF/VCI spectra differ from those in the previous work\cite{Yu2024JCTC} due to the additional constraints now applied to the interaction terms involving cavity modes and molecular vibrations. As expected, an increase in the light-matter coupling factor $g$ results in the rapid splitting of the intricate OH stretch band of 3000-3700 cm$^{-1}$ into two distinct bands, signifying the formation of hybrid vibrational polaritons. The high-frequency spectral band is commonly referred to as the upper polariton (UP) state while the lower-frequency spectral band represents the lower polariton (LP). The spectral separation between UP and LP, known as Rabi splitting, increases with larger light-matter coupling strength. Notably, almost no significant spectral features are observed between UP and LP states, especially when $g$ is substantial. This lack of features does not imply the absence of vibrational states in this region. Instead, many states, known as dark states, exist with frequencies between UP and LP but do not carry infrared intensities.\cite{Yu2024JCTC} Furthermore, it is observed that with strong VSC, the spectral peaks are relatively sharp which is different from the inhomogeneous broadening feature of OH stretches in conventional liquid water. The absence of spectral broadening agrees with previous experimental measurements.\cite{Fukushima2021,Jino2024} This distinctive feature as well as the description of vibrational polaritons serves as additional confirmation of the effectiveness of our fully quantum cav-VSCF/VCI approach. This is further supported by the comparison with recent experimental spectrum in Supplementary Fig. 5, with reasonable Rabi splitting, asymmetry of polaritonic states, and corresponding lineshape. It should be noted again that in VSC experiments, a large number of molecules, i.e. $10^{10}$, collectively couple to the cavity. In our current \ce{(H2O)21}-cavity system, we capture only part of these collective coupling effects. Although our model goes beyond the single-molecule scenario and shows good agreement with the experimental spectrum (Supplementary Fig. 5), the coupling strength factors employed are still significantly larger than those achievable in experiments. Additionally, real cavity systems involve multiple cavity modes, and cavity loss effects must be accounted for. Nevertheless, the qualitative agreement between our theoretical results and experiments provides a reliable platform for exploring the microscopic mechanisms underlying vibrational dynamics. We emphasize that this work considers single cavity mode and no cavity loss effect for the purpose of efficiently investigating how the optical cavity affects the vibrational dynamics. We plan to bridge the gap between current work and more realistic experimental conditions in our future work.

Next, we present how the OH vibrational dynamics are modified by the VSC effect. Fig. \ref{fig2}b showcases examples of vibrational population dynamics after exciting the OH stretch in monomer 1 and monomer 7 under different light-matter coupling strengths. The full results for all 21 monomers are presented in Supplementary Fig. 6-S7. First, as shown in the left panel of Fig. \ref{fig2}b, for certain OH stretches that do not successfully relax when outside the cavity, substantial population decrease is observed under vibrational strong coupling within optical cavity. This cavity-enabled relaxation is generally accelerated with stronger coupling factor $g$. These surprising observations offer direct evidence that VSC introduces new possibilities for vibrational energy transfer in molecular systems, as confirmed in recent experiment.\cite{Xiang2020} For OH stretches undergoing intramolecular vibrational relaxations through Fermi resonance, the inclusion of VSC modifies the rate of population change as shown in the right panel of Fig. \ref{fig2}b. Again, the general trend is that with stronger light-matter coupling, the evolution of vibrational population get apparently speed up.

\begin{figure}[H]
\begin{center}
\includegraphics[width=0.8\textwidth]{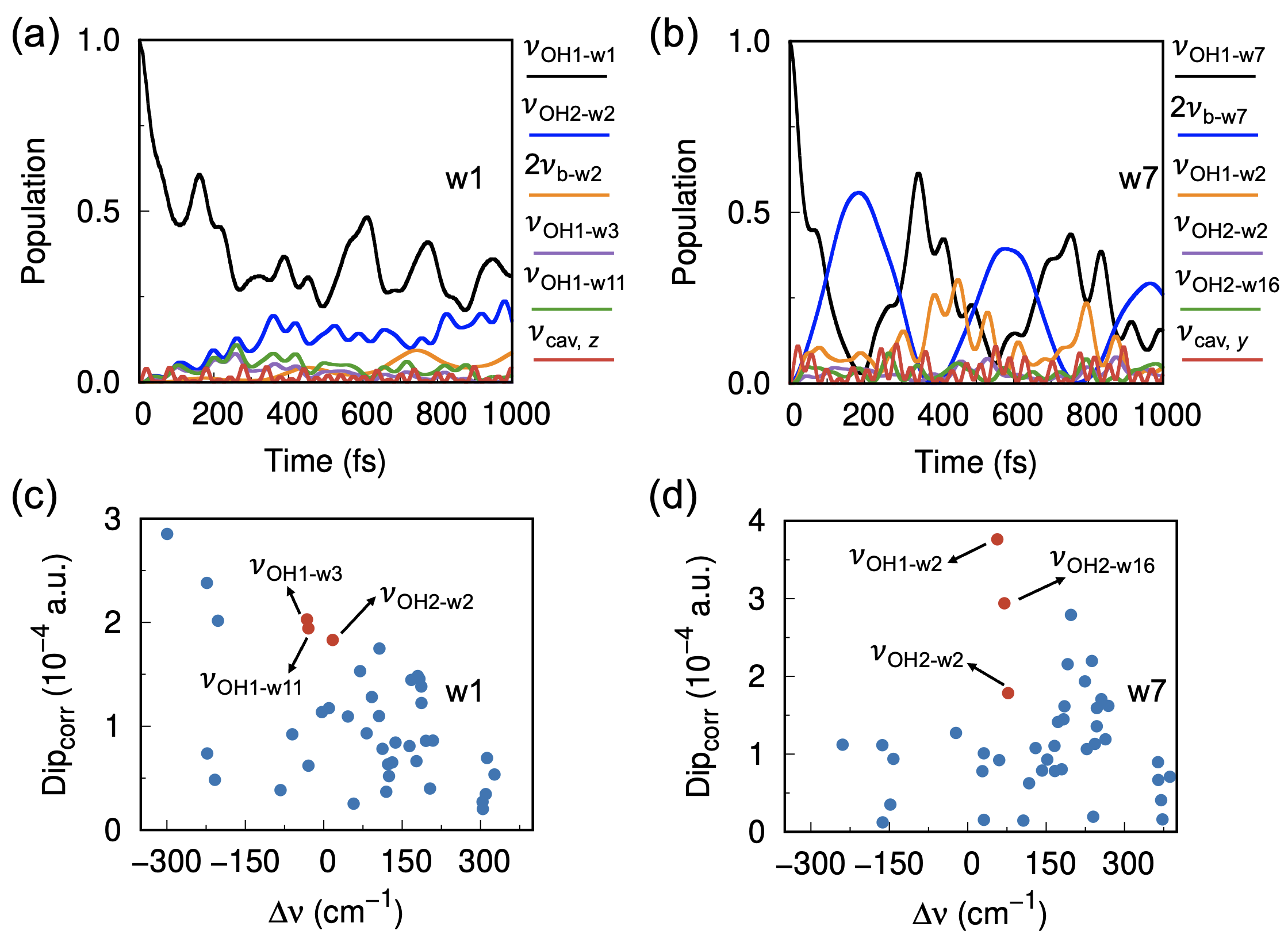}
\end{center}
\caption{\textbf{Vibrational strong coupling effects on OH vibrational energy transfer pathways.} (a-b) Vibrational energy transfer pathways of OH stretch in monomer 1 (w1) and monomer 7 (w7) of \ce{(H2O)21}-cavity system. (c-d) Dipole derivative correlation values and vibrational frequency shift of OH stretches in \ce{(H2O)21} relative to the first OH stretch in monomer 1 and monomer 7 respectively. Red points correspond to primary population receivers observed in (a) and (b) respectively. The dipole derivative correlation value, $\text{Dip}_{\text{corr}}$ is defined in main text. The cavity frequency is set as 3400 cm$^{-1}$ with coupling strength $g=0.006$ (a.u.) for (a)-(d).}
\label{fig3}
\end{figure}

Fig. \ref{fig3}a-b provides detailed vibrational energy transfer pathways when the cavity mode frequency is 3400 cm$^{-1}$ with coupling factor $g=0.006$ a.u.. For the OH stretch in monomer 1, cavity-enabled VEP primarily occurs through direct intermolecular energy transfer among OH groups. The decrease in the population of the excited OH stretch in w1 is accompanied by the significant increase in the population of the OH stretch in monomer 2 (w2). Additionally, notable changes are observed in the populations of the bend overtone in w2 and the OH stretch fundamentals in monomers 3 (w3) and 11 (w11). As listed in Table S1, w2, w3, and w11 belong to the first, second, and third hydration shells of w1, respectively. The strong mixing among these OH stretches indicates the cavity-induced delocalization of the OH stretch as well as efficient intermolecular vibrational energy transfer. It is important to reiterate that our theoretical setup operates under the localization framework where vibrational couplings between vibrational modes of different monomers are ignored, but this constraint is directly broken under VSC. Another surprising finding pertains to the conventionally intramolecular relaxation-dominated OH stretches. As seen in Fig. \ref{fig3}b, for the OH stretch in monomer 7, the dominant VET pathway remains the intramolecular one, where significant population change of the bend overtone is observed. However, the populations of the OH stretches in monomer 2 also increases. As indicated in Supplementary Table 1, monomer 2 is beyond the first hydration shell of monomer 7. Besides the cavity-enabled intermolecular VET between neighbouring molecules, the existence of cavity can provide new energy transfer pathways between remote molecules. The role of the cavity is primarily intermediate, as can be seen in the population change of cavity modes in Fig. \ref{fig3}a-b. More examples of cavity-induced intermolecular VET pathways involving different water monomers are provided in Supplementary Fig. 8. Combing with Supplementary Fig. 6-7, it can be observed that under collective VSC, the vibrational population dynamics of OH stretch, including rate and pathways, in different water monomers vary. This variation is primarily attributed to their specific molecular environments arising from the complex H-bond network. Overall, the increasing rate of energy transfer\cite{Xiong2022science} and the emergence of new intermolecular VET pathways align with recent experimental observations.\cite{Xiang2020}

To delve into the mechanism how VSC modifies the intermolecular VET pathways, we analyzed two key characteristics of all OH stretches. For example, relative to the first OH stretch (3396.77 cm$^{-1}$) in monomer 1, we calculated the spectral shift of other OH stretches across different molecules as $\Delta \nu$ and the dipole derivative correlation defined as \begin{math}\text{Dip}_{\text{corr}}=max(|d\mu_x/dQ_1||d\mu_x/dQ_i|,|d\mu_y/dQ_1||d\mu_y/dQ_i|)\end{math}. These two characteristics describe the similarity between two different OH stretches in terms of their excitation energies and sensitivity along the optical field polarization directions. The scatter plots corresponding to the OH stretches in w1 and w7 are shown in Fig. \ref{fig3}c-d respectively. The primary population receivers during VET process are marked in red. Strikingly, for both w1 and w7, the new and prominent intermolecular VET pathways exhibit the same key features: small vibrational frequency deviations and large dipole derivative correlations. This provides direct evidence of cavity-induced vibrational resonances among OH groups across different water molecules. In addition to frequency similarity, these resonance occur when the dipole derivatives of the OH stretches are both relatively strong and aligned with the cavity's polarization directions. Notably, such resonances can involve both nearby and distant water molecules. For instance, in Fig. \ref{fig3}c, the resonance participants w2 and w11 are located in the first and third hydration shells of w1, respectively. A similar observation is evident in Fig. \ref{fig3}d for monomer 7.

\subsection{Mode-specific mechanism for cavity-modified vibrational dynamics}

\begin{figure}[H]
\begin{center}
\includegraphics[width=0.9\textwidth]{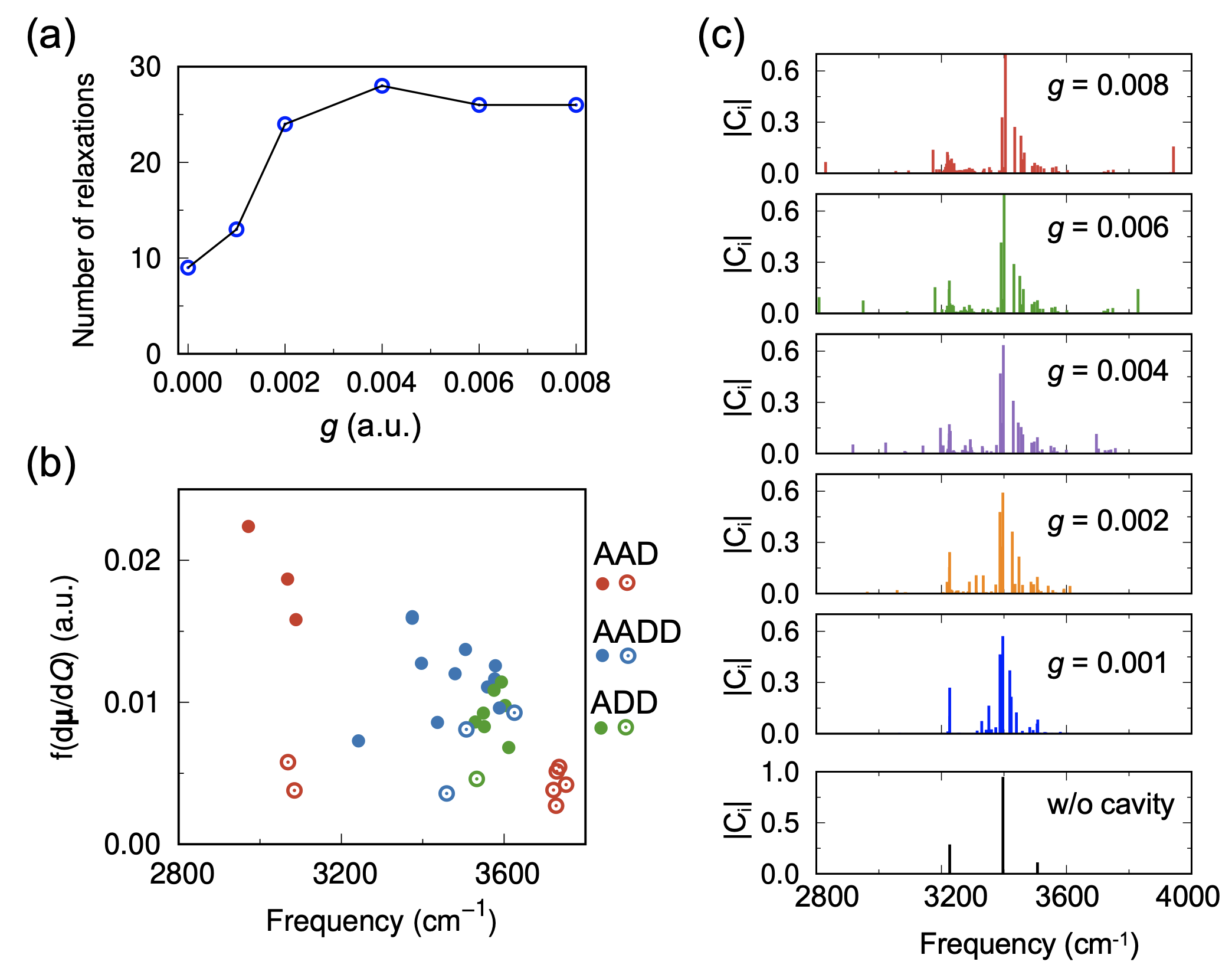}
\end{center}
\caption{\textbf{Mode-specific modification of OH relaxation under VSC} (a) Number of successful relaxations in \ce{(H2O)21}-cavity system as a function of light-matter coupling factors $g$. A successful relaxation is defined that the population of OH stretch can be lower than 0.25 in the time interval of 0-1000 fs. (b) Scattering plot of the magnitude of dipole moment derivative along $x$ and $y$ axis, \begin{math}f(d\boldsymbol{\mu}/dQ)=max(|d\mu_x/dQ|,|d\mu_y/dQ|)\end{math}, of various OH stretches in different types of water molecules which cannot relax outside the cavity. Solid cycles indicate OH stretches can relax inside the cavity while open cycles indicate OH stretches still cannot relax inside the cavity. (c) Stick distributions of VCI coefficients of OH stretch in monomer 1 under different light-matter coupling strengths. The cavity frequency is set as 3400 cm$^{-1}$ for (a)-(c).}
\label{fig4}
\end{figure}

In addition to cavity-induced vibrational resonance, it is essential to explore the general mechanism by which collective VSC influences vibrational dynamics, particularly the common features of vibrations whose relaxation dynamics are altered. Detailed analyses are provided in Fig. \ref{fig4} and Table \ref{tab1}. We begin by examining the number of significant relaxations among the 42 OH stretches, as illustrated in Fig. \ref{fig4}a, with detailed counts listed in Table S3. Outside the cavity, only 9 OH stretches successfully relax via intramolecular vibrational couplings. However, when the \ce{(H2O)21} system is placed inside the cavity, even with a weak light–matter coupling strength of g = 0.001 \text{a.u.}, the number of successful relaxations increases to 13. Under the vibrational strong coupling (VSC) regime, with g = 0.006 \text{a.u.}, this number nearly triples, indicating a dramatic enhancement of relaxation pathways.

\begin{table}[H]
\caption{Number of OH stretches with different vibrational relaxation behaviors. The A/D notation indicates the hydrogen bond acceptor/donor on one water molecule.}
\label{tab1}
\begin{tabular}{c c c c }
\hline\hline
\multirow{2}{*}{No. of OH stretches} &
\multirow{2}{4.5cm}{\centering Successful relaxation \\ outside cavity} &
\multirow{2}{4.5cm}{\centering Successful relaxation \\ inside cavity} &
\multirow{2}{*}{Water type} \\
& & & \\
\hline
5 & Yes & Yes & AADD \\
4 & Yes & Yes & ADD \\
12 & No & Yes &  AADD\\
3 & No & Yes & AAD\\
7 & No & Yes &ADD\\
3 &No & No &  AADD\\
7 &No & No &  AAD\\
1 & No & No & ADD\\
\hline\hline
\end{tabular}
\end{table}

We further classify the 42 OH stretches based on their relaxation behaviors both outside and inside the cavity, as well as their corresponding hydrogen-bonding features. As summarized in Table \ref{tab1}, the 9 OH stretches that successfully relax outside the cavity correspond to H-bonded stretches in AADD and ADD water molecules. Under VSC, the primary effects on these stretches are modifications in their relaxation rates and the emergence of new intermolecular VET pathways, as demonstrated by the example of dynamics of monomer 7 in Fig. \ref{fig2} and Fig. \ref{fig3}. Of the remaining 33 OH stretches that do not relax outside the cavity, 22 achieve relaxation under VSC conditions. These include monomers of various types, such as AADD, AAD, and ADD water molecules. However, 11 OH stretches remain unrelaxed even under VSC conditions, many of which correspond to OH stretches in AAD water molecules, particularly the free OH stretches.

An intuitive explanation for why some OH stretch dynamics are influenced by cavity coupling while others are not lies in the orientation of their transition dipole moments relative to the cavity field polarization. In Fig. \ref{fig4}b, we analyzed the anharmonic vibrational frequencies and dipole derivative information for the 33 OH stretches that do not relax outside the cavity. Similar to the definition above, the dipole derivative for each OH stretch is given by \begin{math}max(|d\mu_x/dQ_i|,|d\mu_y/dQ_i|)\end{math}. As shown by the solid points in Fig. \ref{fig4}b, the OH stretches that exhibit significant changes in their relaxation dynamics under VSC tend to have relatively large dipole derivative magnitudes. These stretches span a broad spectral range of 3000–3600 cm$^{-1}$ and are found in AAD, AADD, and ADD water molecules. In contrast, the 11 OH stretches that remain unrelaxed under VSC generally exhibit smaller dipole derivatives along the cavity’s polarization directions (x and y axes). Notably, these unrelaxed stretches also include H-bonded OH groups with relatively red-shifted frequencies, and nearly half of them correspond to free OH stretches in AAD water molecules.

The cavity-modified vibrational energy transfer rate and pathways can also be attributed to cavity-modified anharmonic vibrational coupling patterns, i.e., the collective couplings among the OH stretches enabled by the VSC. Fig. \ref{fig4}c shows the stick distribution of VCI coefficients of OH stretch in monomer 1 under different light-matter coupling factors $g$. As shown in Equation 6,  the VSCF/VCI approach explicitly and quantitatively provides information on mode couplings among various vibrational states through the corresponding VCI expansion coefficients C$_i$. The magnitude of C$_i$ reflects the extent to which a particular vibrational state contributes to the final VCI eigenstate. Thus, a larger number of VCI eigenstates with nonnegligible contribution from a specific vibration, such as the OH stretch in monomer 1, indicates stronger couplings between this OH stretch and other vibrational states during the formation of hybrid VCI eigenstates. Outside the cavity, the VCI coefficient of OH stretch predominately locates at around 3400 cm$^{-1}$. Due to weak couplings with bend overtone and another high-frequency OH stretch, some components appear at 3240 cm$^{-1}$ and 3507 cm$^{-1}$. When \ce{(H2O)21} is inside the cavity, the molecular coupling patterns are greatly modified. With increasing coupling strength, the VCI coefficient of OH stretches spans an increasing spectral range of 2800-4000 cm$^{-1}$. This is due to the formation of collective polaritonic states and the alignment of OH stretches' transition dipole moment plays an important role in these processes. The cavity-induced direct coupling between OH stretches in different water monomers opens new routes for efficient vibrational relaxations. Even without participation of low-frequency intermolecular modes, the OH stretches show delocalization behavior among waters within and beyond the first hydration shell. 

\begin{figure}[H]
\begin{center}
\includegraphics[width=0.6\textwidth]{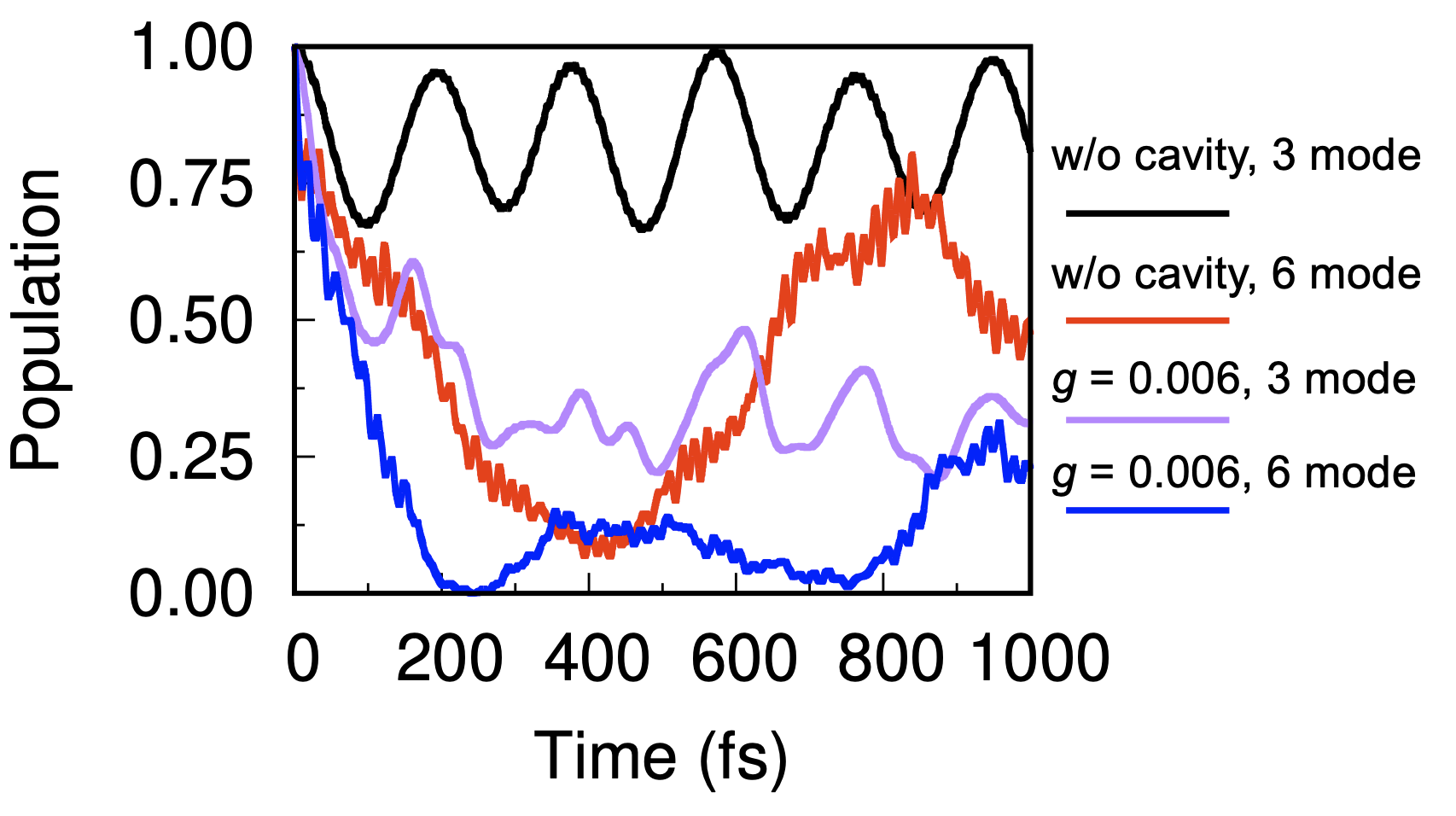}
\end{center}
\caption{Time-dependent population of the OH stretch in monomer 1 (w1) with and without the inclusion of 3 low-frequency intermolecular modes for \ce{(H2O)21} inside and outside the cavity. The cavity frequency is set as 3400 cm$^{-1}$.}
\label{fig5}
\end{figure}

Finally, it is meaningful to investigate the role of low-frequency intermolecular modes for \ce{(H2O)21}-cavity system. We performed additional cav-VSCF/VCI calculation including all intramolecular vibrational modes, the cavity mode, and monomer 1's three low-frequency intermolecular modes. The corresponding dynamics result of OH stretch relaxation is shown in Fig. \ref{fig5}. As discussed in the previous section, the inclusion of three low-frequency modes results in the successful relaxation of the OH stretch for \ce{(H2O)21} outside the cavity. The VSC further enhances the intermolecular relaxation of OH stretches, reinforcing the general finding that efficient vibrational energy transfer among molecules can be achieved through VSC.\cite{Xiang2020}

\section{Discussion}
In this study, we present compelling evidence detailing how the presence of optical cavity influences vibrational energy transfer and dynamics of water. The accuracy of our quantum theoretical approaches is validated in the vibrational spectra and relaxation dynamics of pure water system, aligning well with previous experimental and theoretical calculations. Under the vibrational strong coupling, our cav-VSCF/VCI approach reasonably predicts the polaritonic vibrational spectra, capturing essential spectral features. The VSC effect significantly alters the vibrational population dynamics of OH stretch, impacting both pathways and rates. Notably, VSC breaks the localization framework and directly promotes the delocalization of OH stretch across different water molecules. In addition to the conventional intramolecular relaxation pathway, intermolecular VET pathways are easily realized even without the inclusion of extra intermolecular vibrational motions. This intermolecular energy transfer extends beyond the water's first hydration shell, suggesting the potential for controlling remote molecules through VSC. As to the relaxation rate, the acceleration of vibration energy transfer with stronger light-matter coupling strength is a general trend. All these observations agree with the key findings in recent VSC experiments.\cite{Xiang2020,Xiong2022science}  We also demonstrate that, although exhibiting similar general trends, the specific relaxation dynamics of individual water molecules respond differently to the collective VSC effect in terms of both rate and pathways. 

The primary mechanism underlying cavity-induced vibrational energy transfer is the correlation between the vibrational transition dipole moment vector and the cavity polarization directions. The extent of these dipole correlations plays a crucial role in determining the differences in individual molecules’ vibrational dynamics under VSC. This phenomenon has been explored theoretically in previous studies focusing on model systems .\cite{krupp2024collective, fan2023quantum, mandal2023theoretical} However, to the best of our knowledge, our work provides the first systematic verification of this mechanism through fully quantum simulations of a realistic many-molecules molecular system. Another important mechanism for the emergence of new vibrational energy transfer pathways is cavity-induced vibrational resonance, which involves vibrations in both nearby and remote molecules. The key factors enabling these resonances are the alignment of transition dipole moments with the cavity polarization and the vibrational frequency match between coupled molecules. Our quantum mechanical investigation of mode-specific mechanisms underlying cavity-modified vibrational energy transfer pathways offers new insights into VSC-altered reaction dynamics. These findings can potentially guide the design of appropriate model Hamiltonian systems for analyzing collective VSC dynamics in macroscopic ensembles of molecules within optical cavity.

Our work achieves a fully quantum dynamical simulation of molecule-cavity system on realistic and CCSD(T)-level potential energy surface with collective coupling effect included for the first time. The \ce{(H2O)21} provides a balance between computational efficiency, structural similarity to liquid water, and the incorporation of necessary collective couplings. However, it should be noted again that the realistic cavity system involves significantly more molecules, multiple cavity modes, and the impact of cavity loss should be considered. It remains challenging to quantum-mechanically investigate the effect of cavity-induced modification of vibrational energy transfer as the number of molecules increase. Future calculations will incorporate these effects by extending our approach to condensed-phase system, such as liquid water and dilute HOD in ice.\cite{Hank2014ab} Exploring the effects of vibrational dynamics after exciting polaritonic states and dark states will also be a focus of our future investigations. We hope that our theoretical findings in the current work provide new insights on how the VSC influences the energy transfer and chemical reactivities in molecular systems, as well as stimulating both experimental and theoretical investigations of cavity-induced properties in molecular systems.

\section{Methods}
\subsection{Potential energy and dipole moment surfaces of water}
We employed highly accurate, CCSD(T) level potential energy surface (PES) and MP2 level dipole moment surface (DMS) of water for all the calculations in this work. The PES is taken from our recently developed machine learning potential, q-AQUA.\cite{yu2022q} This potential represents the total energy of $N$-monomer water system \ce{(H2O)n} through standard many body expansion at the 4-body level:
\begin{equation}
\label{pes}
\begin{aligned}
V_{\ce{(H2O)n}}= \sum_{i=1}^N V^{(1)}_{\text{w}_i}+\sum_{i>j}^NV^{(2)}_{\text{w}_i,\text{w}_j}+\sum_{i>j>k}^NV^{(3)}_{\text{w}_i,\text{w}_j,\text{w}_k}+\sum_{i>j>k>l}^NV^{(4)}_{\text{w}_i,\text{w}_j,\text{w}_k,\text{w}_l}
\end{aligned}
\end{equation}
where $V^{(1)}_{\text{w}_i}$ is the spectroscopically accurate 1-body potential for the isolated monomer obtained from Partridge et al.\cite{PS}. $V^{(2)}_{\text{w}_i,\text{w}_j}, V^{(3)}_{\text{w}_i,\text{w}_j,\text{w}_k}$, and $V^{(4)}_{\text{w}_i,\text{w}_j,\text{w}_k,\text{w}_l}$ are the intrinsic 2-, 3-, 4-body interactions among water molecules. The 2-body potential is from a machine learning fit using 71,892 CCSD(T)/CBS 2-b energies. The 3-body and 4-body potentials were fitted based on 45,332 BSSE corrected CCSD(T)-F12a/aVTZ and 3692 CCSD(T)-F12a/haTZ data respectively. The accuracy of q-AQUA has been systematically verified in structures, vibrational spectra, and dynamics tests across gas-phase water clusters and condensed phase liquid water.\cite{yu2022q}

The dipole moment of water is taken from our previously developed WHBB DMS\cite{liu2012quantum} which also follows a many-body representation:
\begin{equation}
\label{d2b}
\begin{aligned}
\mu_{\ce{(H2O)n}}= \sum_{i=1}^N \mu^{(1)}_{\text{w}_i}+\sum_{i>j}^N\mu^{(2)}_{\text{w}_i,\text{w}_j}
\end{aligned}
\end{equation}
where the 1-body water dipole $\mu^{(1)}_{\text{w}_i}$, is taken from a highly accurate DMS for isolated monomer developed by Lodi et al.\cite{h2odip} The 2-body dipole, $\mu^{(2)}_{\text{w}_i,\text{w}_j}$, is from a machine learning fit to roughly 30,000 MP2/aVTZ dipole moment data.\cite{liu2012quantum} The final DMS for water has also been tested extensively in the IR spectral calculations of water clusters, liquid water, and ice. More details of the PES and DMS of water dimer are referred to the q-AQUA water potential\cite{yu2022q} and WHBB water DMS.\cite{liu2012quantum} 

\subsection{cav-VSCF/VCI approach}
We recently developed cavity vibrational self-consistent field/virtual state configuration interaction (cav-VSCF/VCI) approach to calculate polaritonic vibrational spectra of molecular system in an optical cavity.\cite{Yu2022,Yu2024JCTC} This approach employs the Pauli-Fierz Hamiltonian \cite{Rubio_2017_jctc,mandal2023theoretical} to describe the system of $N$ molecules with $N_Q$ mass-scaled rectilinear normal modes \textbf{Q} in an optical cavity of $N_\text{C}$ cavity modes:
\begin{equation}
\label{H_tot}
\begin{aligned}
\hat{H}_{\text{QED}}&=-\frac{1}{2}\sum_{i}^{N_{\text{Q}}}\frac{\partial^2}{\partial Q_i^2}+ \sum_k^{N_\text{C}}\frac{1}{2}\hat{p}_k^2+V_{\text{eff}}(\textbf{Q},\textbf{q})
\end{aligned}
\end{equation}
where $V_{\text{eff}}(\textbf{Q},\textbf{q})$ is the effective potential energy of the molecule-cavity system, $V_{\text{eff}}(\textbf{Q},\textbf{q})=V(\textbf{Q})+\sum_k^{N_\text{C}}\frac{1}{2}\omega_k^2(\hat{q}_k+\sqrt{\frac{2}{\omega_k^3}}g\hat{\boldsymbol{\mu}}\cdot\textbf{e}_k)^2$. $\hat{p}_k$ and $\hat{q}_k$ are the momentum and position operators for cavity mode $k$. $\hat{\boldsymbol{\mu}}$ is the dipole moment vector for the molecular system. $\textbf{e}_k$ is the polarization vector of the field for the cavity mode $k$. Each cavity mode $k$ is assumed to have frequency $\omega_k$ and polarization vector $\textbf{e}_k$.  $g$ is the light-matter coupling strength factor which is related to cavity mode frequency, permittivity $\epsilon_0$, and effective cavity volume  $\tilde{V}$, such that $g= \sqrt{\omega_k/(2\epsilon_0\tilde{V})}$.

In analogous to the conventional VSCF/VCI approach, our cav-VSCF/VCI method expresses the hybrid nuclear-cavity vibrational wavefunction on the basis of corresponding VSCF states:
\begin{equation}
\begin{aligned}
\Psi(\textbf{Q},\textbf{q})&=\sum_m C_m\Phi_m(\textbf{Q},\textbf{q})\\
&=\sum_mC_m\prod_{i=1}^{N_{\text{Q}}}\phi_i^{m}(Q_i)\prod_{i=1}^{N_\text{C}}\chi_i^{m}(q_i) \\
\end{aligned}
\end{equation}
where $\Phi_m(\textbf{Q},\textbf{q})$ is the quantum VSCF state as direct product of $N_{\text{Q}}$ molecular modes,$\phi_i^{m}(Q_i)$, and $N_{\text{C}}$ cavity modes, $\chi_i^{m}(q_i)$. The VCI combination coefficients $C_m$ are determined by diagonalizing the VCI Hamiltonian matrix. More details of the cav-VSCF/VCI method are referred to our previous work.\cite{Yu2022,Yu2024JCTC}

A series of technical strategies have recently been  implemented to enable large-scale cav-VSCF/VCI calculations on the large cavity-molecules system, i.e., \ce{(H2O)21}.\cite{Yu2024JCTC} However, due to non-zero interaction terms between modes from different molecules, these implementations cannot provide an absolutely localized scheme among molecules. Thus it could not be directly employed to study how VSC modifies the localization scheme of the molecular system. In this work, we address this problem by imposing an additional constraint on the effective potential,$V_{\text{eff}}(\textbf{Q},\textbf{q})$. Under the $n-$mode representation framework, $V_{\text{eff}}(\textbf{Q},\textbf{q})$ is expressed as:
\begin{equation}
\label{nmode}
\begin{aligned}
& V_{\text{eff}}(\textbf{Q},\textbf{q})=\sum_iV_i^{(1)}(Q_i)+\sum_iV_i^{(1)}(q_i)\\
&+\sum_{i,j}V_{ij}^{(2)}(Q_i,Q_j)
+\sum_{i,j}V_{ij}^{(2)}(Q_i,q_j)\\
&+\sum_{i,j}V_{ij}^{(2)}(q_i,q_j)+\sum_{i,j,k}V_{ijk}^{(3)}(Q_i,Q_j,Q_k) + \cdots\\
\end{aligned}
\end{equation}
where $V_i^{(1)}(Q_i)$ and $V_i^{(1)}(q_i)$ are the one-mode potentials for the molecular normal mode and the cavity mode, respectively. $V_{ij}^{(2)}(Q_i,Q_j)$, $V_{ij}^{(2)}(q_i,q_j)$, $V_{ij}^{(2)}(Q_i,q_j)$, etc. are the intrinsic 2-mode potentials for pairs of molecular normal modes, pairs of cavity modes, and pairs composed of a molecular normal mode and a cavity mode, etc. 

To reduce the computational cost and employ the local monomer approximation strategy,\cite{wang2011ab,yu2017high} our previous work ignores significant number of coupling terms between intramolecular vibrations from different molecules.\cite{Yu2024JCTC} For example $V_{ij}^{(2)}(Q_i,Q_j)$ and $V_{ijk}^{(3)}(Q_i,Q_j,Q_k)$ are set as 0 if one of the normal modes is from a different molecule. Additionally, in this work, to ensure absolutely zero  couplings between intramolecular vibrations of different molecules, we impose a further constraint such that $V_{ijk}^{(3)}(q_i,Q_j,Q_k)$ is also set as 0 if $Q_j$ and $Q_k$ are from different molecules. It should be noted that to fully describe the vibrational dynamics of OH stretches, the local monomer approximation employed in this work is not sufficient as the OH stretch excitations can delocalize over its hydration shells, $\sim$ 15 water molecules.\cite{Tokmakoff2013,kraemer2008temperature} However, together with previous implementations, this new framework provides a platform to investigate how vibrational strong coupling affects  molecular coupling patterns and localization behavior.

The new cav-VSCF/VCI framework has been implemented in our in-house version of MULTIMODE\cite{multimode,yu2022multimode} with an interface to the q-AQUA PES and WHBB DMS. Note, both PES and DMS are full dimensional which are used to obtain the effective potentials $V_{\text{eff}}$ at different nuclear configurations. Instead of frequently used Condon approximation, in our VSCF/VCI framework, the full dimensional DMS is further employed to calculate the transition dipole moment for IR intensity calculation using VCI wavefunctions.\cite{burcl2003infrared} In our new cav-VSCF/VCI calculations of \ce{(H2O)21}-cavity system, the \ce{(H2O)21} is places inside an optical cavity oriented along the $x$ axis. A single cavity mode with two polarization directions ($y$ and $z$) is considered. In most calculations, three intramolecular modes (2 OH stretches and 1 HOH bend) of each water monomer are included. Thus, at least 63 localized normal modes of  \ce{(H2O)21} are considered. A 3-mode representation (3MR) of the effective potential, $V_{\text{eff}}$, and a 2-mode representation (2MR) of the dipole moment were used. 

\subsection{Quantum wavepacket dynamics simulation}
The quantum wavepacket dynamics were conducted based on the cav-VSCF/VCI wavefunctions. Similar approach has been applied in vibrational relaxation dynamics calculations of dilute HOD in ice.\cite{Hank2014ab} For simplicity, we rewrite the $i_\text{th}$ cav-VSCF/VCI wavefunction, $\Psi_i$ as: 
\begin{equation}
\begin{aligned}
\Psi_i&=\sum_j^{N_{\text{b}}} C_{ij} \Phi_j\\
\end{aligned}
\end{equation}
where $N_{\text{b}}$ is the total number of VSCF basis functions $\Phi_j$, $C_{ij}$ is the VCI expansion coefficient. 
Now, we use the non-stationary VSCF state to represent the initial wavepacket of OH stretch excitation, such that
\begin{equation}
\begin{aligned}
\psi_k(0)&= \Phi_k=\sum_j^{N_{\text{b}}}C_{kj}^{-1}\Psi_j\\\
&=\sum_j^{N_{\text{b}}}C_{kj}^T\Psi_j=\sum_j^{N_{\text{b}}}C_{jk}\Psi_j
\end{aligned}
\end{equation}
where the orthogonal property of the eigenvector matrix $C$ is applied.
Thus, the time-dependent wavepacket is expressed by
\begin{equation}
\begin{aligned}
\psi_k(t)&=\sum_j^{N_{\text{b}}}C_{jk}\Psi_je^{-iE_jt/\hbar}
\end{aligned}
\end{equation}
The auto-correlation function can be directly calculated as
\begin{equation}
\begin{aligned}
Corr_k(t)=\langle \psi_k(0) | \psi_k(t) \rangle=\sum_j^{N_{\text{b}}}C_{jk}^2e^{-iE_jt/\hbar}
\end{aligned}
\end{equation}
As seen, $Corr_k(t)^2$ is the population change of the initial wavepacket $\psi_k(0)$ and time-dependent population of another state $\Phi_l$ can be calculated as $Corr_k^l(t)^2$ where $Corr_k^l(t)$ is given by:
\begin{equation}
\begin{aligned}
Corr_k^l(t) = \langle \Phi_l(0) | \psi_k(t) \rangle = \sum_j^{N_{\text{b}}}C_{jl}C_{jk}e^{-iE_jt/\hbar}
\end{aligned}
\end{equation}
We propagated the wavepacket dynamics for 1000 fs for each set of excited OH stretch. The population change for the initial excited OH stretch and the population of other states are recorded for further analyses. Note, our wavepacket dynamics are based on the VCI eigenstates and coefficients obtained at the stage of VSCF/VCI computations. Although this approach differs from conventional time-dependent quantum wavepacket dynamics, it is well-suited for simulating time-dependent population dynamics and has been successfully employed to analyze the vibrational relaxation pathways of dilute HOD in Ice Ih.\cite{Hank2014ab}

\bibliography{refs}

\section{Data availability}
The data generated and used in this study are available at upon request to the authors.

\section{Code availability}
The codes used in this study are available at upon request to the authors.

\section{Acknowledgements}
Q.Y. acknowledges the support from National Natural Science Foundation of China (22473030). J.M.B. acknowledges the support from NASA grant (80NSSC22K1167). D.H.Z. acknowledges the support from the National Natural Science Foundation of China (grant no. 22288201).

\section{Author contributions}
Q.Y. conceived the project, performed calculations and analyzed the data. Q.Y., D.H.Z., and J.M.B. discussed the results and wrote the manuscript.

\section{Competing interests}
The authors declare no competing interests.

\section{Additional information}
Supplementary information is available.

\end{document}